\documentclass[prb,aps,twocolumn,showpacs,floats,amssymb,amsmath,amsfonts]{revtex4}

\usepackage{graphicx}
\usepackage{bm}
\renewcommand{\vec}{\bm}

\begin{document}

\title{
Modeling scanning tunneling spectra of Bi$_2$Sr$_2$CaCu$_2$O$_{8+\delta}$
}

\author{
B. W. Hoogenboom,\cite{address} C. Berthod, M. Peter,\cite{note} and
\O. Fischer
}

\affiliation{
DPMC, Universit\'e de Gen\`eve, 24 Quai Ernest-Ansermet,
1211 Gen\`eve 4, Switzerland
}

\author{A. A. Kordyuk}

\affiliation{
Institute for Solid State Research, IFW Dresden, P.O. Box 270016,
D-01171 Dresden, Germany\\
Institute of Metal Physics of the National Academy of Sciences of Ukraine,
03142 Kyiv, Ukraine
}

\date{\today}

\begin{abstract}

Recent angle-resolved photoemission and neutron scattering data have provided
new ingredients for the interpretation of scanning tunneling spectra on
Bi$_2$Sr$_2$CaCu$_2$O$_{8+\delta}$. We analyze the low-temperature tunneling
spectra, from oxygen overdoped to underdoped samples, including details about
the bilayer splitting and the neutron resonance peak. Two van Hove
singularities are identified: the first is integrated in the coherence peaks,
the second is heavily broadened at higher binding energy. The shape of the
tunneling spectra suggests a strong coupling of the quasiparticles with a
collective mode, and a comparison with photoemission shows that the scattering
rate in tunneling is an order of magnitude smaller than in ARPES. Finally, the
theoretical spectra calculated with an isotropic tunneling matrix element are
in better agreement with the experimental data than those obtained with
anisotropic matrix elements.

\end{abstract}

\pacs{74.50.$+$r, 73.40.Gk, 74.72.$-$h, 74.72.Hs}
% 74.50.+r (tunnel/s.c.), 73.40.Gk (tunnel), 74.72.-h (HTS), 74.72.Hs (BSCCO)
\maketitle

\section{Introduction}

Scanning tunneling spectroscopy (STS) is a powerful tool to study the
electronic properties of solids. Its remarkable energy and spatial resolution
makes it particularly well suited for materials characterized by small energy
and short length scales, like the cuprate high-$T_c$ superconductors (HTS).
Among the HTS, the bilayer compound Bi$_2$Sr$_2$CaCu$_2$O$_{8+\delta}$ (BSCCO)
has often been studied by STS, because it cleaves easily and offers an
atomically flat BiO surface. It is possible to tunnel through the insulating
BiO and SrO surface layers into the CuO$_2$ plane, where all the exciting
properties of the cuprates are believed to reside. Many important results on
the nature of the superconducting, normal, and mixed state of BSCCO have been
obtained using STS.\cite{Renner:1998a,Miyakawa:1999,Hudson:1999,%
Zasadzinski:2001,Pan:2001,Hoogenboom:2001,Hoffman:2002,Misra:2002}

In the superconducting state, the main feature of the differential tunneling
conductance ($dI/dV$) spectrum is the quasiparticle excitation gap, which has
been observed in BSCCO and studied as a function of doping and
temperature.\cite{Renner:1998a,Miyakawa:1999,Kaneko:1998,Matsuura:1998,%
Matsuda:1999} The presence of excitations within the superconducting gap,
linearly increasing with energy around $V=0$, indicates that the order
parameter has nodes, and presumably $d_{x^2-y^2}$ symmetry. Other
characteristics of the BSCCO spectra are the celebrated dip-hump structure at
energies larger than the excitation gap, an asymmetry between electron and hole
tunneling, and coherence peaks with considerably more spectral weight than
predicted by BCS theory. None of these three characteristics has been met with
an explanation that is commonly agreed upon.

Under certain assumptions,\cite{Mahan:1983,Harrison:1961} and considering the
nearly two-dimensional nature of BSCCO, one finds that the shape of the STS
spectrum is determined by only three ingredients: the bare electron dispersion
$\varepsilon_{\vec{k}}$ in the CuO$_2$ plane, the self-energy
$\Sigma(\vec{k},\omega)$ which embodies all electronic interactions, and a
tunneling matrix element $T_{\vec{k}}$ which couples the electronic states of
momentum $\vec{k}$ at the sample surface with the metallic tip. In the
particular case where $T_{\vec{k}}$ is a constant, the STS spectrum directly
relates to the quasiparticle density of states (DOS) in the CuO$_2$ plane.
Behind this apparently simple statement lies a real difficulty to disentangle
the contribution of each of the above ingredients. Various features in the STS
spectra were thus attributed either to van Hove singularities in the
band-structure\cite{Bok:1997,Wei:1998} or to self-energy
effects.\cite{Hirsch:1999,Cren:2000} Furthermore, the general shape of the
low-temperature spectrum was considered suggestive of an anisotropic matrix
element.\cite{Wei:1998,Kouznetsov:1996,Yusof:1998,Franz:1999}

The situation has recently become even more complicated with the observation of
a clear bilayer splitting by angle-resolved photoemission
(ARPES).\cite{Feng:2001,Chuang:2001,Kordyuk:2002} The two CuO$_2$ layers in the
BSCCO unit cell give rise to two non-degenerate bands close to the Fermi
energy. As a consequence, there are two bands --- instead of one, as assumed
previously --- which can contribute to the $dI/dV$ spectra. Using ARPES, the
shape of these bands has been determined in the normal state of underdoped and
overdoped BSCCO.\cite{Kordyuk:2002b}

In fact, the correct interpretation of STS spectra relies upon a
realistic modeling of the data, and different models can lead to
opposite conclusions. Moreover, the modeling critically depends on
details of the band structure\cite{Kordyuk:2002b} and of the spin
excitation spectrum,\cite{Fong:1999,He:2001} which have become
available only very recently.

In this study we compare the predictions of several models to the
STS spectra measured at low temperature on BSCCO samples with
different oxygen dopings levels.\cite{Renner:1998a} Our
calculations take into account the bilayer splitting and are based
on the band structures determined in
Ref.~\onlinecite{Kordyuk:2002b}. We assume a pure $d_{x^2-y^2}$
symmetry of the superconducting gap and consider three different
models for the self-energy: a conventional BCS model; a
phenomenological model proposed to fit the ARPES
data,\cite{Kordyuk:2002b} thus allowing a direct comparison
between photoemission and tunneling; and a model which describes
the coupling of quasiparticles to a collective
mode.\cite{Eschrig:2000} We also compare the effects of isotropic
and anisotropic tunneling matrix elements.

\section{Modeling the tunneling spectra}

In the tunneling-Hamiltonian formalism,\cite{Mahan:1983} the sample and tip are
coupled by a matrix element $M_{\vec{k}\vec{q}}$ which represents the overlap
of the electronic states on both sides of the tunnel junction. The resulting
differential conductance at bias voltage $V$ is given by
   \begin{equation}\label{eq:dIdV}
     \frac{dI}{dV}\propto-\int d\omega\sum_{\vec{k},n}
       |T_{\vec{k}}|^2 A_n(\vec{k},\omega) f'(\omega-eV)
   \end{equation}
where $f$ is the Fermi function and $A_n$ is the spectral function
in the sample. The sign convention is such that, at zero
temperature, negative energies $\omega$ and $eV$ correspond to
occupied states, and positive energies to unoccupied states. The
new matrix element appearing in Eq.~(\ref{eq:dIdV}) is
$|T_{\vec{k}}|^2 = \sum_{\vec{q}}|M_{\vec{k}\vec{q}}|^2
A_\text{tip}(\vec{q},\omega)$. Using a tip with a featureless DOS,
we can assume that it is energy independent. According to
Ref.~\onlinecite{Harrison:1961} the dependence of $T_{\vec{k}}$ on
$k_z$ is cancelled by the band dispersion along $k_z$ ($z$ is the
tunneling direction). In Eq.~(\ref{eq:dIdV}), $n$ refers to the
two bands resulting from the bilayer splitting as discussed below.
We assume that the $k_{xy}$ dependence of $T_{\vec{k}}$ is the
same for both bands; this then leads to an $n$-independent matrix
element. $A_n$ is related to the electron dispersion and
self-energy through
   \begin{equation}\label{eq:A}
     A(\vec{k},\omega)=-\frac{1}{\pi}\,\text{Im}\,\left[\frac{1}
     {\omega+i\Gamma-\varepsilon_{\vec{k}}-\Sigma(\vec{k},\omega)}\right].
   \end{equation}
The band index is omitted for simplicity. The lifetime broadening $\Gamma$ is
introduced here for computational convenience and is set to $\Gamma=1$~meV in
all of our calculations.

To account for the bilayer splitting, we use the band structures determined in
Ref.~\onlinecite{Kordyuk:2002b} for the anti-bonding (A, plus sign) and bonding
(B, minus sign) bands:
   \begin{eqnarray}\label{eq:band}
     \nonumber
     \varepsilon_{\vec{k}}^{\text{(A,\,B)}}
     &=&-2\,t(\cos k_x+\cos k_y)+4\,t'\cos k_x\cos k_y\\
     \nonumber
     &&-2\,t''(\cos2k_x+\cos2k_y)\\
     &&\pm {\textstyle\frac{1}{4}}\,t_\perp(\cos k_x-\cos k_y)^2+
       \Delta\varepsilon.
   \end{eqnarray}
The interlayer coupling is described by $t_\perp$; the maximum energy splitting
between the A and B bands is $2t_\perp$ and coincides with the van Hove
singularities at the $(\pi,\,0)$ point in the Brillouin zone. The tight-binding
parameters inferred from ARPES for overdoped (OD69.0K) and underdoped (UD77.0K)
BSCCO are given in Table~\ref{TB_par}. These parameters deviate slightly from
those reported in Ref.~\onlinecite{Kordyuk:2002b}. $t_\perp$ was determined in
Ref.~\onlinecite{Kordyuk:2002b} for the OD69.0K sample. The determination of
$t_\perp$ for the UD77.0K sample is complicated by the strong influence of the
pseudo- and superconducting gaps on the photoemission spectra. For that reason,
UD77.0K was assumed to have the same $t_\perp$ as OD69.0K in
Ref.~\onlinecite{Kordyuk:2002b}. In this work, we have estimated $t_\perp$ for
UD77.0K from a careful comparison between leading edge gaps of the A and B
bands.\cite{Borisenko:2002} The remaining parameters were modified accordingly.
This leads to a better agreement with the tunneling data. From the parameters
of OD69.0K and UD77.0K, we make a linear interpolation with doping to obtain
the parameters appropriate for the samples studied by STS
(Table~\ref{TB_par}).\cite{extra_pol}

\begin{table}[b!]
\caption{\label{TB_par}
Tight-binding parameters for the conduction bands of BSCCO. The parameters for
OD69.0K and UD77.0K follow from fits to ARPES data, and in the other cases from
a linear interpolation with doping. All numbers are in eV.
}
\begin{ruledtabular}
\begin{tabular}{cccccc}
Sample & $t$ & $t'$ & $t''$ & $t_\perp$ & $\Delta\varepsilon$\\ \hline
OD69.0K & 0.40 & 0.090 & 0.045 & 0.082 & 0.431 \\
UD77.0K & 0.39 & 0.076 & 0.034 & 0.097 & 0.304 \\ \hline
OD56.0K & 0.40 & 0.092 & 0.047 & 0.081 & 0.449 \\
OD74.3K & 0.40 & 0.089 & 0.044 & 0.084 & 0.422 \\
OP92.2K & 0.39 & 0.082 & 0.039 & 0.090 & 0.361 \\
UD83.0K & 0.39 & 0.077 & 0.035 & 0.096 & 0.317 \\
\end{tabular}
\end{ruledtabular}
\end{table}

The first model we consider is the conventional BCS model. In this case the
self-energy reads:
   \begin{equation}\label{eq:BCS}
     \Sigma^{(1)}(\vec{k},\omega)
       =\frac{|\Delta_{\vec{k}}|^2}{\omega+i\Gamma+\varepsilon_{\vec{k}}},
   \end{equation}
where the gap has $d_{x^2-y^2}$ symmetry:
$\Delta_{\vec{k}}=\frac{1}{2}\,\Delta_0(\cos k_x-\cos k_y)$. The only free
parameter, $\Delta_0$, is adjusted to the experimental data.

In the second model, we follow an attempt to fit ARPES
measurements\cite{Kordyuk:2002b} by complementing the BCS model with a
phenomenological $\vec{k}$-independent self-energy $\Sigma_0(\omega)$. At low
energy, $\Sigma_0(\omega)$ has a marginal Fermi liquid form:
   \begin{equation}\label{eq:MFL1}
     \Sigma_0(\omega)
       =-\lambda\omega-i\sqrt{(\alpha_0\omega)^2+(\beta_0 k_{\text{B}}T)^2}.
   \end{equation}
Good agreement with ARPES was found with the parameters $\lambda=1$,
$\alpha_0=2$, and $\beta_0=4$.\cite{Kordyuk:2002b,Kordyuk:2002a,Lanzara:2001,%
Johnson:2001,Gromko:2002} The complete self-energy including the
superconducting gap is
   \begin{equation}\label{eq:MFL2}
     \Sigma^{(2)}(\vec{k},\omega)
       =\Sigma_0(\omega)+\frac{|\Delta_{\vec{k}}|^2}
       {\omega+i\Gamma+\varepsilon_{\vec{k}}-\Sigma_0(\omega)}.
   \end{equation}
Eq.~(\ref{eq:MFL2}) can be recast in a form similar to Eq.~(\ref{eq:BCS}), with
the bare dispersion $\varepsilon_{\vec{k}}$ and gap $\Delta_{\vec{k}}$
renormalized to $c\,\varepsilon_{\vec{k}}$ and $c\,\Delta_{\vec{k}}$,
respectively, and the lifetime broadening $\Gamma$ replaced by
$\Gamma-c\,\text{Im}\,\Sigma_0(\omega)$ with $c=(1+\lambda)^{-1}$. We will use
Eq.~(\ref{eq:MFL2}) to check if the self-energy inferred from ARPES is
compatible with the STS data.

Finally, the third model we consider accounts for the possible interaction of
the quasiparticles with a collective mode. Such a mode was indeed observed in
BSCCO by neutron scattering, with momentum centered at
$(\pi,\,\pi)$.\cite{Fong:1999} It appears at an energy $\Omega \approx
5.4\,k_{\text{B}}T_c$,\cite{He:2001} and is characterized by a correlation
length $\xi\lesssim2$ in lattice units (we will take $\xi=2$ in our
simulations). The interaction of this mode with the electrons was described
theoretically in Ref.~\onlinecite{Eschrig:2000}, and involves a coupling
constant $g$ for which we choose the value $g=0.65$~eV. The self-energy
entering Eq.~(\ref{eq:A}) is
   \begin{equation}\label{eq:mode}
     \Sigma^{(3)}(\vec{k},\omega)=\Sigma_{11}(\vec{k},\omega)+
       \frac{|\Delta_{\vec{k}}+\Sigma_{12}(\vec{k},\omega)|^2}
       {\omega+i\Gamma+\varepsilon_{\vec{k}}-\Sigma_{22}(\vec{k},\omega)}.
   \end{equation}
The components $\Sigma_{ij}$ are a convolution of the bare BCS propagator (in
Nambu representation) with the spin susceptibility. The latter is represented
by a simple analytical function which approximates the neutron measurements. We
refer the reader to Ref.~\onlinecite{Eschrig:2000} for further details.

Very little is known about the actual $\vec{k}$-dependence of the tunneling
matrix element, although band calculations suggest that it is anisotropic with
a shape corresponding to the dispersion of the bilayer
splitting,\cite{Andersen:1995} i.e. $(\cos k_x-\cos k_y)^2$. Such a matrix
element would prohibit tunneling into the nodal direction $(\pi,\,\pi)$ and
would highlight the region $(\pi,\,0)$ of the van Hove singularities. We will
consider the two limiting cases of a completely isotropic $T_s=T_0$ and
anisotropic $T_d=T_0(\cos k_x-\cos k_y)$ matrix element, as well as admixtures
of the form $|T_{\vec{k}}|^2 = \alpha|T_s|^2+(1-\alpha)|T_d|^2$.

For each model we calculate the differential tunneling conductance using
Eq.~(\ref{eq:dIdV}) and a $1024\times1024$ mesh of $\vec{k}$ points. The
temperature is set to 4.2~K. The experimental energy resolution is simulated by
filtering the data with a Gaussian of width 1~meV. Finally, unless stated
otherwise, the calculated spectra are normalized to the total spectral weight
of the experimental data over the energy range from $-300$ to $+300$~meV.

\section{Results and discussion}

\subsection{BCS model}\label{sect:BCS}

\begin{figure}[tb!]
\includegraphics[width=75mm]{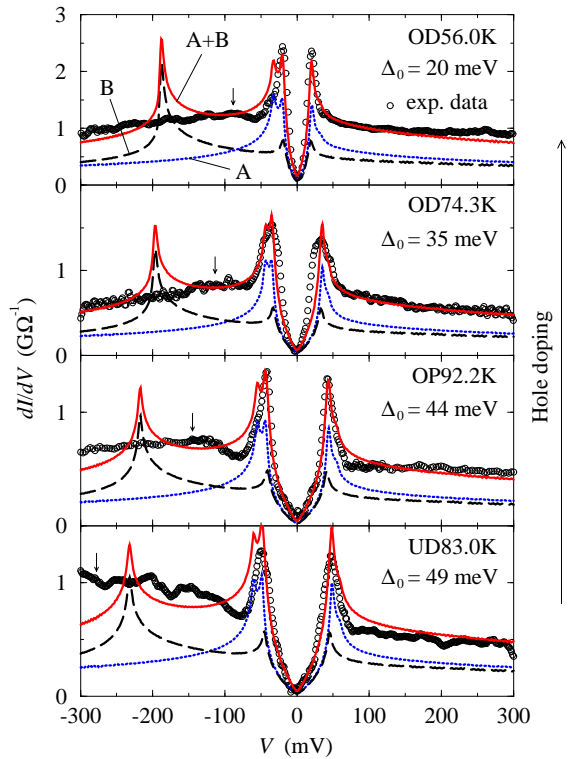}
\caption{\label{Fig1} Tunneling spectra in the conventional BCS
model, with a $d$-wave superconducting gap and an isotropic matrix
element. The contributions of the A and B bands are shown
separately, and their sum should be compared to the experimental
data (circles). The arrows roughly indicate the maximum of the
broad feature in the background (see text). }
\end{figure}

To compare the calculated spectra with experimental results over a wide doping
range, we use low-temperature (4.2~K) data of Renner {\it et
al.}\cite{Renner:1998a} The results of the BCS model, with an isotropic matrix
element, are shown in Fig.~\ref{Fig1}. The agreement at subgap energies is
good, except for OD74.3K where the spectral weight is somewhat underestimated.
The (calculated and experimental) spectra show the V-shape at zero bias typical
for a $d$-wave order parameter. At higher energies, however, the model misses
the dip-hump feature, which is most pronounced in the optimally doped and
underdoped spectra at $-60$ to $-100$~meV, and the model generally presents too
sharp structures. In particular, the van Hove singularities of the A and B
bands show up unrenormalized in the spectra while no such singularities exist
in the experimental curves. None of these discrepancies can be reduced by the
inclusion of an anisotropic matrix element, which would suppress spectral
weight below the gap energy and raise the van Hove singularities even more.
Nevertheless, with this first approach we can draw some conclusions which
remain valid for the more sophisticated models discussed below.

From the difference between the A and B bands, it becomes clear that most of
the weight of the coherence peaks in the tunneling spectra is related to the
van Hove singularity of the A band, which lies close to the Fermi level. The
presence of the van Hove singularity in the coherence peaks explains their
unusual height.\cite{Wei:1998} Note, however, that without the B band the peaks
would become too high (with respect to the background) to fit the experimental
data. The van Hove singularity also explains why the peak at negative bias is
generally a bit higher than the one at positive bias.

In general, we observe that the van Hove singularity of the A band remains
close to the Fermi level (integrated in the coherence peaks) for all doping
levels. The van Hove singularity of the B band, however, moves to higher
binding energy with underdoping. This shift is related to the combined effects
of increasing gap and increasing $t_{\perp}$ with underdoping. It is in
qualitative agreement with the behavior observed in the background of the
experimental data:\cite{Renner:1998a,Miyakawa:1999} on going from overdoped to
underdoped, the background of the spectra becomes more asymmetric. One can
distinguish a very broad feature moving away from the Fermi level with
underdoping. In fact, we have shown that this background can be fitted very
well by a broadened van Hove singularity at energies consistent with the doping
level, in a rigid single-band picture.\cite{Hoogenboom:2002PhD} These energies
are indicated by arrows in Fig.~\ref{Fig1}. In the present calculation,
however, the van Hove peaks are located at different energies.

Summarizing this discussion, the tunneling spectra are consistent with a sharp
van Hove singularity just below the Fermi level, and a very broad van Hove
singularity moving to higher binding energy with underdoping. Apart from the
dip-hump feature, the questions left open by the BCS model are the precise
energy shift of the van Hove singularity of the bonding band, as well as the
mechanism of its broadening.

\subsection{Comparison of tunneling and ARPES}

The absence of a sharp van Hove singularity of the B band in the tunneling
spectra is a clear indication that one should go beyond the bare BCS DOS to
explain all features in the spectra. In principle, it should be possible to
take spectral functions directly from ARPES data and sum them to compare to the
tunneling results. However, this is complicated by the strong influence of
matrix elements on the photoemission spectra (see e.g.
Ref.~\onlinecite{Kordyuk:2002a}) and of the background in the ARPES data. We
therefore rely on the phenomenological model of the spectral function at low
energy described above [with self-energy Eq.~(\ref{eq:MFL2})]. This description
crudely includes the effects of correlation in the marginal Fermi liquid
self-energy $\Sigma_0(\omega)$. With such an approach we can first directly
compare tunneling and ARPES, and also study the ability of the marginal Fermi
liquid model to account for tunneling data.

\begin{figure}[tb!]
\includegraphics[width=75mm]{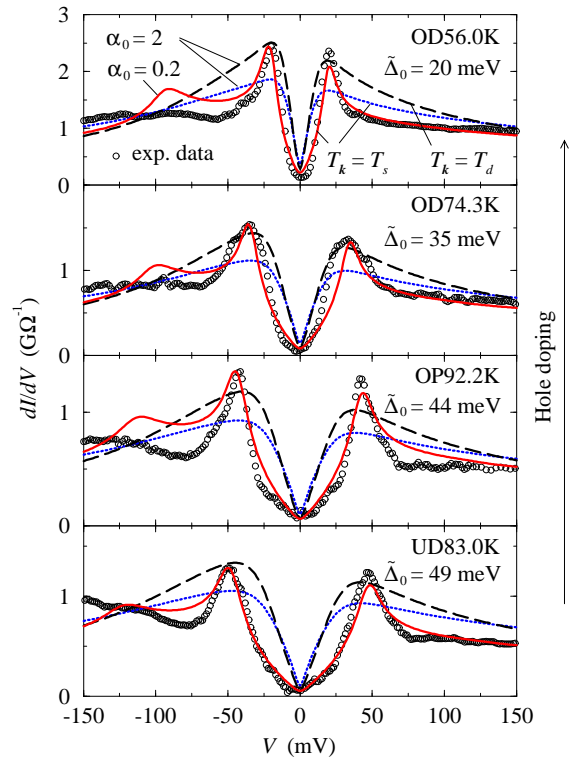}
\caption{\label{Fig2}
Tunneling spectra calculated using the phenomenological self-energy
Eq.~(\ref{eq:MFL2}) and compared to the experimental data (circles). Spectra
are shown for $\alpha_0=2$ with the isotropic (dotted curve) and anisotropic
(dashed curve) matrix element, the latter highlighting the ($\pi,\,0$) region
of the Brillouin zone, and for $\alpha_0=0.2$ with the isotropic matrix
element. The two curves for $\alpha_0=2$ are normalized to the spectral weight
between $-300$ and $+300$~meV, while the curve for $\alpha=0.2$ is normalized
to the peak height. $\tilde{\Delta}_0=(1+\lambda)^{-1}\Delta_0$ is the
renormalized gap value. All spectra are for $\lambda=1$ and $\beta_0=4$.
}
\end{figure}

The ARPES data could be fitted using Eq.~(\ref{eq:MFL2}) with $\alpha_0=2$. In
Fig.~\ref{Fig2} we compare the calculations for $\alpha_0=2$ to experimental
tunneling data. Note that the energy scale is a factor 2 smaller than in
Figs.~\ref{Fig1} and \ref{Fig3}. As can be seen, the calculated curves (dotted
lines) do not fit the experimental data at all. In particular, the V-shape of
the spectra at low energy is too narrow. One might expect that the excess
spectral weight in the gap can be suppressed by an anisotropic matrix element.
However, as shown by the dashed lines, this is not the case. The reason is that
the low-energy spectral weight is dominated by the marginal Fermi liquid
self-energy, and not by the gap function.

Surprisingly, a good fit can be obtained by an order of magnitude reduction of
$\alpha_0$, to $\alpha_0=0.2$, assuming an isotropic matrix element. This
difference in $\Sigma''_0=\text{Im}\,\Sigma_0$ for ARPES and tunneling
suggests, not for the first time,\cite{Franz:1998a} that the scattering rate in
ARPES is considerably higher than in tunneling. The explanation for this
remarkable difference goes beyond our current understanding of ARPES and
tunneling.

The van Hove singularity of the B band is shifted closer to the Fermi level as
a result of band renormalization by a factor $(1+\lambda)^{-1}$. Furthermore,
the energy dependence of $\Sigma''_0$ leads to a broadening which is more
important at higher energies. At negative bias this results in a coherence peak
followed by the broadened van Hove singularity of the B band. However, the
energy shift of this van Hove singularity as a function of doping is too small
compared to the broad feature in the background (see Section~\ref{sect:BCS}).
Of course one can question the validity of the assumption
$\text{Re}\,\Sigma_0=-\lambda\omega$ at energies above
$\sim50$~meV,\cite{Johnson:2001} meaning that the van Hove singularity of the B
band is at too low binding energy in our simulations. Taking this into account
would probably lead to a better agreement with the broad maximum in the
background, shifting away from the Fermi level with underdoping. However, this
behavior is inconsistent with the sharpness of the dip when the van Hove
singularity is at much higher binding energy than the dip (as is specifically
the case for the optimally and underdoped samples).

\subsection{Coupling to a collective mode}

\begin{figure}[b!]
\includegraphics[width=75mm]{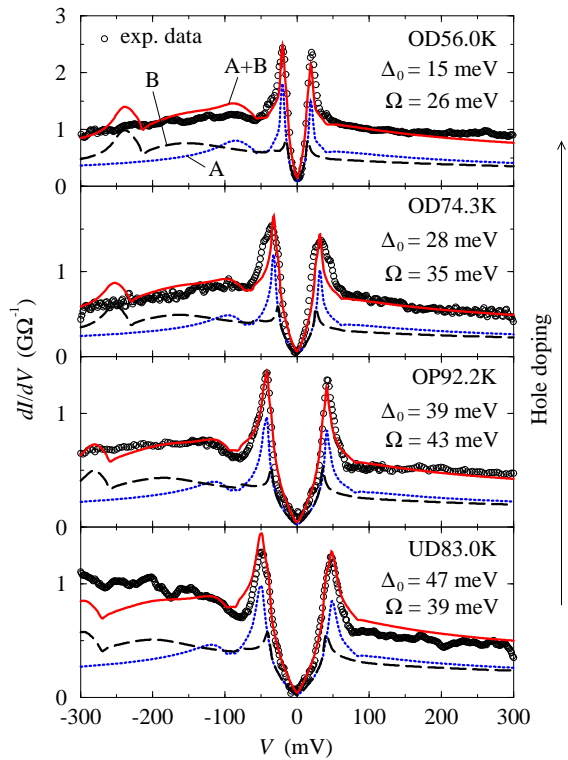}
\caption{\label{Fig3}
Tunneling spectra including interaction with a bosonic mode at wave vector
($\pi,\,\pi$) and energy $\Omega=5.4\,k_{\text{B}}T_c$. The coupling constant
and correlation length are $g=0.65$~eV and $\xi=2$, respectively. The
superconducting gap has $d$-wave symmetry and the tunneling matrix element is
isotropic. The contributions of the A and B bands are shown separately, and
their sum should be compared to the experimental data.
}
\end{figure}

So far, it has not been established whether the coupling of
quasiparticles to the collective mode observed in neutron
scattering experiments, is sufficient to have a sizable influence
on the spectral functions\cite{Abanov:2002} or not.\cite{Kee:2002}
In the following analysis we assume it is, and verify to which
extent the tunneling spectra are consistent with neutron
scattering data on the $(\pi,\,\pi)$ mode.

In this model [Eq.~(\ref{eq:mode})], the main effect of the resonant mode is to
enhance the imaginary part $\Sigma''$ of the self energy between the energies
$\varepsilon_1=-\Omega-\Delta_0$ and
$\varepsilon_2=-\Omega-\sqrt{\varepsilon_{\text{vHs}}^2+\Delta_0^2}$, where
$\varepsilon_{\text{vHs}}$ is the band energy at the ($\pi,\,0$)
point.\cite{Eschrig:2000} A similar effect, although much smaller, exists at
positive bias. The numerical results are displayed in Fig.~\ref{Fig3}. Focusing
first on curves A, which represent the DOS of the A band, we notice the sharp
coherence peaks. The energy of the latter is a combination of the gap value and
the van Hove singularity which, as in the previous models, lies very close to
the gap edges, and contributes much to the total spectral weight of the
coherence peaks. Furthermore, it is crucial for the creation of the dip-hump
feature, below the coherence peak at negative bias. The width of the dip
corresponds to the energy interval between $\varepsilon_1$ and $\varepsilon_2$
where $\Sigma''$ is enhanced. Below the dip, there is a hump resulting from the
scattering out of the above energy interval. The mode energy deferred from
neutron scattering \cite{Fong:1999,He:2001} results in a reasonable agreement
with the dip in the tunneling spectra (see also
Ref.~\onlinecite{Zasadzinski:2001}). However, we caution that its shape is more
dependent on the precise values of $\Delta_0$ and $\varepsilon_{\text{vHs}}$
than on the mode energy itself. Furthermore, for the optimally and underdoped
samples the calculated dip is at slightly higher binding energy than in the
experimental data, suggesting that the mode energy is smaller than determined
from neutron scattering. Finally, the depth of the dip in the experimental data
indicates an increasing coupling constant $g$ with underdoping.

Let us now turn to the DOS of the B band. Here the coherence peaks are small
and occur at $\Delta_0$. The energy interval between $\varepsilon_1$ and
$\varepsilon_2$ is much larger than for the A band, and below $\varepsilon_2$
we see again a hump structure. The van Hove peak of the B band now appears
inside the dip energy range as a very broad maximum. Comparing the curves at
different dopings, we find that the broad maximum shifts away from the Fermi
level with underdoping. Though this shift is smaller than suggested by
experiment, it is in qualitative agreement with experiment. The only
disagreement occurs at higher binding energies, where the hump coming from the
B band is not observed in experiment. This hump is probably smeared out by the
continuum of spin excitations at higher energies,\cite{Eschrig:2002} not taken
into account here.

\begin{figure}[b!]
\includegraphics[width=75mm]{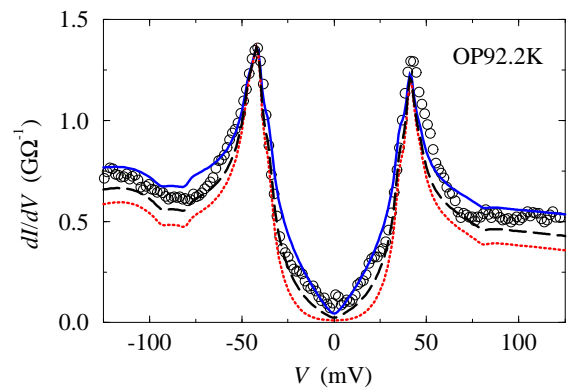}
\caption{\label{Fig4}
Comparison of the STS spectra including the coupling to the ($\pi,\,\pi$) mode,
calculated with different matrix elements. $\Delta_0 = 39$~meV, as in
Fig.~\ref{Fig3}. The continued line corresponds to the isotropic case
$T_{\vec{k}}=T_s$, the dotted line to the anisotropic case $T_{\vec{k}}=T_d$,
and the dashed line to a partly anisotropic matrix element
$|T_{\vec{k}}|^2=0.4|T_s|^2+0.6|T_d|^2$. The spectra have been normalized to
the peak height at $-44$~meV.
}
\end{figure}

In Fig.~\ref{Fig4} we compare experimental data to spectra calculated using
different matrix elements. It is clear that the experimental spectrum at low
bias shows the V-shape typical for a $d$-wave superconductor, and {\em not\/}
the U-shape expected for tunneling with a completely anisotropic matrix
element. Though some anisotropy is not to be excluded, the tunneling spectra
are thus more consistent with isotropic than with anisotropic tunneling matrix
elements that suppress all states along the diagonals of the Brillouin zone.
This conclusion is independent of the exact model chosen here (it can also be
witnessed from, for example, the bare DOS in Fig.~\ref{Fig1}), and only
dependent on the assumption of $d$-wave symmetry of the order parameter and a
small imaginary part of the self-energy at low energy.

\section{Summary}

We have modeled scanning tunneling spectra of BSCCO including $d$-wave BCS
superconductivity, band dispersions based on recent ARPES data (with bilayer
splitting), and isotropic as well as anisotropic tunneling matrix elements. In
addition to this, we have compared tunneling spectra to ARPES data via a
phenomenological marginal Fermi liquid approach. In a third model, we have
taken into account coupling of quasiparticles to a collective mode with
momentum ($\pi,\pi$). All numerical results have been compared to experimental
data over a large doping range.

The simple $d$-wave BCS model reproduces several general characteristics of the
experimental tunneling spectra, but fails to account for salient features like
the dip-hump structure and the absence of a sharp van Hove singularity from the
bonding band. The comparison between tunneling and ARPES suggests a much longer
lifetime in tunneling experiments than in ARPES: using the marginal Fermi
liquid approach, the shape of the coherence peaks in tunneling spectra is
correctly reproduced, but only if the lifetime is taken an order of magnitude
larger than inferred from ARPES data, an observation that remains to be
explained. This model, specified for low energy, does not lead to a
satisfactory description of the dip-hump structure and the asymmetric
background as a function of doping. Important improvement is obtained by
including the interaction of quasiparticles with a collective mode. With
parameters inferred from neutron scattering experiments, reasonable agreement
with the tunneling spectra is found, though the energy of the mode may be
slightly overestimated for the optimally and underdoped samples.

In general, the tunneling spectra are consistent with the presence of a sharp
van Hove singularity (of the antibonding band) integrated in the coherence
peaks, and a broad van Hove singularity (of the bonding band) in the
background, shifting away from the Fermi level with underdoping. These van Hove
singularities can be held responsible for the asymmetry in tunneling spectra
with respect to zero bias. The broadening of the bonding band van Hove
singularity can be largely attributed to the collective mode. This mode also
leads to a dip-hump feature in the spectra, which is more pronounced in the
underdoped than in the overdoped samples. The exact shape and depth, however,
also depend on van Hove singularity of the antibonding band and on the
superconducting gap.

Finally, the shape of the spectra at low-bias voltage is indicative of an
isotropic tunneling matrix element. We conclude that the tunneling matrix
element does not have a strong dependence on the (in-plane) wave vector.
Tunneling spectroscopy therefore probes states along the whole Fermi surface,
including the diagonals of the Brillouin zone.

\begin{acknowledgments}

We acknowledge A. A. Manuel for programming assistance, Ch. Renner for sharing
his long experience in tunneling spectroscopy, and G. A. Sawatzky for his
repeatedly questioning about tunneling matrix elements. This work has been
supported by the Swiss National Science Foundation.

\end{acknowledgments}

%\bibliographystyle{prsty}
%\bibliography{STS_simulation}

\end{document}